\documentclass{aa}  

\usepackage{amsmath}
\usepackage{txfonts}
\usepackage{booktabs}
\usepackage{array}
\usepackage{graphicx}
\usepackage{natbib}
\usepackage{upgreek}
\usepackage{color}
\usepackage{xcolor}
\usepackage{wasysym}
\usepackage[normalem]{ulem}
\usepackage[colorlinks,linkcolor=blue,anchorcolor=blue,citecolor=blue,breaklinks=true]{hyperref}
\usepackage{tablefootnote}
\bibpunct{(}{)}{;}{a}{}{,} 
\usepackage{linenoaa}

\begin{document} 

   \title{The cosmological lithium problem}

   \subtitle{}

   \author{Oswaldo D. Miranda
          \thanks{oswaldo.miranda@inpe.br}
          }

   \institute{Instituto Nacional de Pesquisas Espaciais,
              Av. dos Astronautas 1758 -- Jardim da Granja \\
                S\~ao Jos\'e dos Campos, SP 12227--010, Brazil}
             
   \titlerunning{The cosmological lithium problem}

   \authorrunning{Miranda}          

   \date{Received ; accepted }

 
  \abstract
   {The discrepancy between the predictions of primordial nucleosynthesis and the observed lithium abundance in Spite plateau stars has been attributed either to a challenge to the standard model of nucleosynthesis or to stellar processes occurring after the stars formed. To understand the origin of this discrepancy, it is crucial to link the cosmic star formation rate with a chemical enrichment model that incorporates the yields of both Population (Pop) III and II stars. It is within this framework that the evolution of lithium can be determined.}  
   {The primary goal is to demonstrate that there is no discrepancy between the predictions of primordial nucleosynthesis and the observed lithium abundance.}   
   {By combining a standard chemical evolution model with the hierarchical structure formation scenario, it is possible to determine the lithium abundance as a function of $[\mathrm{Fe/H}]$. The model's results are compared with observational data from extremely metal-poor stars, Spite plateau stars, Gaia-Enceladus sources, the Small Magellanic Cloud, lithium abundances in Solar System meteorites, and two extremely iron-poor stars: J0023+0307 and SMSS J0313--6708.}  
   {The Spite plateau is naturally established in the range $-8.0 \lesssim [\mathrm{Fe/H}] \lesssim -2.0$ with $^{7}\mathrm{Li/H}$ $\sim 1.81 \times 10^{-10}$. We find that J0023+0307 could have formed $\sim 4.4 \times 10^{5} - 1.3 \times 10^{6}$ years after the explosion of the first Pop III star in the Universe, whereas for SMSS J0313--6708 this event would have occurred $\sim 2.2 \times 10^{5} - 4.4 \times 10^{5}$ years later.}
   {The Spite plateau serves as an observational signature of the formation of Pop III stars. The abundances observed in J0023+0307 and SMSS J0313--6708 are consistent with Pop III progenitor stars in the mass range $10-100 M_{\odot}$. However, if some high-redshift star formation occurs within subhalo-like structures, the contribution of stars in the mass range $140-260 M_{\odot}$ to the formation of the extended Spite plateau cannot be ruled out.}
   
   \keywords{cosmology: theory --- cosmology: observations --- primordial nucleosynthesis --- first stars --- stars: Population III --- stars: Population II}
   
\maketitle

\nolinenumbers

\section{Introduction}
\label{sec1}

The standard big bang nucleosynthesis (BBN) framework connects cosmology with the standard model of particle physics, describing the formation of light elements ($\mathrm{D}$, $^{3}\mathrm{He}$, $^{4}\mathrm{He}$, $^{7}\mathrm{Li}$, and $^{7}\mathrm{Be}$) during the first minutes after the big bang. The only free parameter in  standard BBN is the baryon-to-photon ratio ($\eta$), which is now best determined through measurements of the cosmic microwave background (CMB) radiation by instruments aboard the \textit{Wilkinson Microwave Anisotropy Probe} (\textit{WMAP}) and \textit{Planck} satellites \citep{r1,r2,r3}. In particular, CMB measurements of baryon density serve as inputs for primordial nucleosynthesis calculations, enabling predictions of elemental abundances that can be directly compared with observational data (see, e.g., \citealp{r4} and the references therein).

The combination of BBN and CMB predictions shows excellent agreement with high-redshift observational data for $\mathrm{D}$ and $^{4}\mathrm{He}$. However, the same is not true for $^{7}\mathrm{Li}$: BBN+CMB predictions suggest a primordial abundance of $(^{7}\mathrm{Li/H})_{\mathrm{BBN+CMB}}= (4.94 \pm 0.72)\times 10^{-10}$ \citep{r5,r6}, whereas metal-poor near-main-sequence stars in the Spite plateau indicate an observed value\footnote{Where $({^7}\mathrm{Li/H}) \equiv N({^7}\mathrm{Li})/N(\mathrm{H})$.} of $({^7}\mathrm{Li/H})_{\text{obs}} = (1.6 \pm 0.3) \times 10^{-10}$ (\citealp{r6} and references therein). This discrepancy, with the theoretical value exceeding the observed one by a factor of $\sim 3-4$, is known as the cosmological lithium problem.

Many studies have attempted to address the lithium problem, proposing various hypotheses to explain the discrepancy between theoretical predictions and observations. One straightforward explanation is that primordial lithium was depleted through stellar processes post-formation. Stellar models that incorporate diffusion and lithium transport mechanisms have been developed to quantify lithium abundances in old stars, partially explaining the observed levels (see, e.g., \citealp{r8,r9,r10}). Additional processes such as rotational mixing, turbulence, and magnetic field effects can further reduce surface lithium over time, particularly in low-mass stars, helping explain Spite plateau lithium levels despite variations in metallicity and stellar temperature among Population (Pop) II stars \citep{r11}. In particular, a compilation of key studies on lithium abundances in metal-poor near-main-sequence stars over the past few decades can be found in \citet{r7}.

An alternative explanation posits that primordial nucleosynthesis predictions require modification due to physics beyond the standard model. The existence of exotic particles or modifications in baryon--photon interactions in the early Universe could have altered lithium production during primordial nucleosynthesis (\citealp{r4}, see also \citealp{r12} for a recent review).

Similarly, supersymmetry  theories introduce particles that could have influenced lithium evolution by acting as catalysts for nuclear reactions or interfering with beryllium decay, thereby affecting lithium production \citep{r13}. Dark matter (DM) interactions have also been explored, with hypotheses suggesting that DM particles decayed during or shortly after primordial nucleosynthesis, releasing energy and modifying nuclear reaction rates relevant to lithium abundances \citep{r14}. Another possibility, motivated by models involving dynamic scalar fields, such as Guts or dilaton theories, suggests that the fine-structure constant was larger than the current value by $2-20\,\mathrm{ppm}$ during the BBN period (see \citealp{r56}; \citealp{r57}; \citealp{r58}). While these alternative scenarios are theoretically well motivated, they introduce additional complexities when confronted with observational data.

The Spite plateau was first identified in low-metallicity Galactic stars \citep{r15}, which are considered good indicators of the primordial lithium abundance due to their minimal contamination from later nuclear processes. The Sagittarius (Sgr) dwarf galaxy was among the first extragalactic systems studied for Spite plateau lithium \citep{r16,r17}. Low-metallicity stars in Sgr exhibit lithium abundances consistent with the Spite plateau observed in the Milky Way, though slightly lower, potentially reflecting differences in the star formation history and chemical evolution of these two galaxies.

The Sculptor dwarf galaxy is another extragalactic system in which the lithium abundance in metal-poor stars has been studied, with observations confirming the presence of a Spite plateau in its oldest stars \citep{r18,r19}. More recently, \citet{r20} analyzed stars belonging to the Gaia-Enceladus galaxy and found a Spite plateau similar to that of the Milky Way. The detection of a lithium plateau across different environments, particularly in nearby dwarf galaxies, suggests that primordial lithium production was largely uniform across galaxies. Small variations observed in some systems may offer valuable insights into evolutionary processes influencing lithium in distinct galactic environments.

In this study we incorporated lithium into the semi-analytical model proposed by \citet{r21}, originally developed to explore the cosmological evolution of metallicity in the Universe. This model accounts for contributions from Pop III and Pop II stars to the synthesis of 11 elements: Fe, Si, Zn, Ni, P, Mg, Al, S, C, N, and O.
 
In Sect. \ref{sec2} we review the model from \citet{r21}, hereinafter referred to as the CMW model, which couples the cosmic star formation rate (CSFR) with chemical evolution equations, and discuss how lithium is integrated into this framework. In Sect. \ref{sec3} we examine the role of Pop III and Pop II stars in lithium evolution, emphasizing the significance of nova systems and Galactic cosmic rays (GCRs) for $^{7}\mathrm{Li}$ production when metallicity\footnote{Where $\mathrm{[Fe/H]} \equiv \log_{10}\,\mathrm{[(Fe/H)_{obs}/\mathrm{(Fe/H)}_{\odot}]}$.} exceeds $[\mathrm{Fe/H}] > -1.0$. Conclusions are presented in Sect. \ref{final}.

\section{Methodology}
\label{sec2}

The cosmological model used by CMW assumes that DM halos form at redshift $z \sim 20$ with masses on the order of $10^{6}M_{\odot}$. More massive structures form later as the redshift decreases. Using the Press-Schechter formalism \citep{r22}, the number of DM halos per comoving volume at a given redshift $z$ and with masses in the interval $[M,M+dm]$ can be determined (see in particular Sect. 2.1 of CMW).

Once formed, DM halos generate potential wells that draw primordial gas into their interiors. In this way, the chemical elements synthesized by primordial nucleosynthesis (such as hydrogen, helium, and lithium) are gradually incorporated into the halos through an infall process. The physical parameter associated with this process is the baryon accretion rate, given by

\begin{equation}
\label{baryon}
a_\mathrm{b}(t) = \Omega_{0,\mathrm{b}}\,\rho_{0,\mathrm{c}} \left(\frac{dt}{dz}\right)^{-1} \left| \frac{df_\mathrm{b}}{dz} \right|,\end{equation}
where $\Omega_{0,\mathrm{b}}$ is the baryonic density parameter at $z = 0$, $\rho_{0,\mathrm{c}} = 3H_{0}^{2}/8\pi G$ is the critical density of the Universe ($H_{0} = 100\, h \, \mathrm{km} \, \mathrm{s}^{-1} \mathrm{Mpc}^{-1}$ is the Hubble parameter at the present time), and $f_\mathrm{b} (z)$ is the fraction of primordial baryons incorporated into the halos as determined from the Press-Schechter formalism. The cosmological framework determining the function $a_\mathrm{b}(t)$ is similar to that used by other authors (e.g., \citealp{r23,r24,r25,r26,r27}).

As gas gradually accumulates inside virialized halos via $a_\mathrm{b}(t)$, conditions are created for stars to begin forming within these structures. Using Schmidt's law \citep{r28}, which characterizes the conversion of gas into stars, and Salpeter's law \citep{r29}, which characterizes the initial mass function (IMF), the CSFR represented by $\dot{\rho}_{\star}(z)$ can be determined (see Sect. 2.2 of \citealp{r21}). This characterizes the cycle associated with star formation, which is primarily related to the removal of gas to form stars, the return of chemically processed gas by stars to the environment, and the formation of stellar remnants at the end of the stars' life.

The mechanisms regulating $\dot\rho_{\star}(z)$ are described by the star formation efficiency $<\varepsilon_\star(z)>$ and the star formation timescale $\tau_\mathrm{s}(z)$, which represents the gas depletion timescale. As discussed in CMW, at high redshifts, the $\tau_\mathrm{s}(z)$ function is dominated by the atomic gas depletion timescale, while for $z < 1.0$, $\tau_\mathrm{s}(z)$ is dominated by the molecular depletion time. On the other hand, the $<\varepsilon_\star(z)>$ function exhibits nearly constant behavior for $z > 3.5$, with values between $\sim 0.15$ and $0.40$, depending on the IMF exponent. For $z < 3.5$, the star formation efficiency gradually decreases with redshift, reaching values close to $0.01-0.02$ at $z = 0$.

Additionally, the cosmological scenario in this study incorporates the results of \citet{r26}, in which the authors demonstrated the connection between the CSFR and local (Galactic) star formation (LSF). In particular, \citet{r26} identified that turbulence has a dual role in star formation: it induces the star formation efficiency to remain high and almost constant until a critical Mach number ($M_\mathrm{crit}$) is reached, after which it induces a rapid decline in the star formation efficiency.

The unified model of \citet{r26} consistently derives Larson's  first law \citep{r30} associated with the velocity dispersion $(<V_{\mathrm{rms}}>)$ in local star-forming regions. Finally, the most important characteristic of \citet{r26}, relevant to this work, is that part of the star formation taking place inside a halo with mass $\sim 10^{6} M_{\odot}$ occurs in substructures (or subhalos) with masses similar to those of globular clusters. It is within this cosmological framework that the analysis of the lithium abundance is performed.

\subsection{Lithium yields of Pop III and Pop II stars}
\label{sub21}

Once star formation begins at $z \sim 20$, conditions are created for the chemical enrichment of the Universe. The first stars to form have zero metallicity (Pop III) and gradually enrich the environment. As discussed in CMW, upon reaching the transition metallicity $Z = 10^{-6}$, conditions are created for the formation of the first generation of Pop II stars. As a result, Pop III stars are no longer formed, but due to the mass that governs the lifetime $\tau_\mathrm{m}\,(Z=0)$, some zero-metallicity stars can coexist with Pop II stars with a metallicity of $10^{-6}$, as long as $\tau_\mathrm{m}\,(Z=0) > t_{1}$ (where $t_{1}$ is the moment in time when the metallicity of the Universe reaches $10^{-6}$).

When the metallicity of the Universe reaches $Z = 10^{-4}$, the second generation of Pop II stars is formed. This generation can coexist with some of the stars from previous generations (Pop III and Pop II of the $Z=10^{-6}$ branch), as long as the conditions $\tau_\mathrm{m} \,(Z=0) > t_{2}$ and $\tau_\mathrm{m}\,(Z=10^{-6}) > t_{2}$ are satisfied (where $t_{2}$ is the moment in time when the metallicity of the Universe reaches $10^{-4}$). This process repeats itself at each metallicity transition in the Universe determined by the values $10^{-3}$, $4\times 10^{-3}$, $8\times 10^{-3}$, and $2\times 10^{-2}$. Thus, the stars of Pop II and Pop III are situated in different classes of evolutionary trajectories dictated by the metallicity of the Universe and the lifetime of the stars ($\tau_\mathrm{m}$), which is itself a function of the metallicity (see CMW and references therein). The reference $t_{0}$ is determined by the age of the Universe at redshift $z=20$.

Tables \ref{tab01} and \ref{tab02} show the ranges of stellar masses and metallicities from which the yields were chosen. The Pop II branches shown in Table \ref{tab02} were chosen to provide the best combination of masses and metallicity ranges, enabling the construction of a consistent chemical model encompassing transitions across stellar populations from metallicity $10^{-6}$ to $2\times 10^{-2}$. Additionally, the Pop II stars in Table \ref{tab02} share the same parametric basis --- MONSTAR code for stellar evolution, OPAL opacities, and compatible mass loss models (see \citealp{r21} and references therein) --- ensuring that the chemical scenario is built on a common physical foundation.
\begin{table}[!ht]
\begin{center}
\centering
\caption{Masses selected for Pop III chemical yields.}
\label{tab01}
\begin{tabular}{lcll}
\hline
\hline

Model & \multicolumn{1}{l}{CL08} & HW10   & HW02 \\
\hline \hline
Metallicity ($Z$) & \multicolumn{3}{c}{Mass ($M_{\odot}$)} \\ \hline
\multicolumn{1}{c}{$0$}   & 0.85--3.0  & \multicolumn{1}{c}{10--100} & \multicolumn{1}{c}{140--260}  \\ \hline
\end{tabular}
\end{center}
\tablefoot{CL08: \citet{r31}, HW10: \citet{r33}, and  HW02: \citet{r32}.}
\end{table}

\begin{table}[!ht]
\begin{center}
\centering
\caption{
Masses and metallicities selected for Pop II chemical yields.}
\label{tab02}
\begin{tabular}{ccccc}
\hline
\hline

\multicolumn{1}{l}{Model}  & K10   & D14a       & D14b   & CL04     \\ \hline \hline
\multicolumn{1}{l}{Metallicity ($Z$)} & \multicolumn{4}{c}{Mass ($M_{\odot}$)}  \\ \hline
$10^{-6}$    & --     & --         & --                 & 13--35  \\
$10^{-4}$    & 1--6 & --           & 6.5--7.5 & 13--35  \\
$10^{-3}$    & --     & --         & 6.5--7.5 & 13--35  \\
$4 \times 10^{-3}$   & 1--6 & 6.5--8.0 & --             & --        \\
$8 \times 10^{-3}$   & 1--6 & 6.5--8.5 & --             & --        \\
$2 \times 10^{-2}$   & 1--6 & 7.0--9.0 & --             & 13--35        \\ \hline
\end{tabular}
\end{center} 
\tablefoot{K10: \citet{r34}, D14a: \citet{r35}, D14b: \citet{r36}, and CL04: \citet{r37}.}
\end{table}

\subsection{Contribution of primordial lithium to the structures in formation}
\label{sub22}

The function $a_\mathrm{b}(t)$ given in Eq. (\ref{baryon}) supplies primordial gas to the structures formed, and it can be written as

\begin{equation}
\label{ali}
a_\mathrm{b}(t) = a_\mathrm{b-H}(t) + a_\mathrm{b-He}(t) + a_\mathrm{b-Li}(t),
\end{equation}
where $a_\mathrm{b-H}(t)$ is the accretion rate of primordial hydrogen, $a_\mathrm{b-He}(t)$ accounts for the accretion of primordial helium, and $a_\mathrm{b-Li}(t)$ is the corresponding accretion rate of primordial lithium. It is important to highlight that primordial gas is continuously injected into the star-forming environment through the function $a_\mathrm{b}(t)$. This term, in turn, depends on the halo mass function, involving halos of different masses that decouple from the Universe's expansion at different redshifts. Once injected into the star-forming environment, part of the primordial gas is used to form stars, which is quantified, at each time t, by the parameter $<\varepsilon_{\star}(z)>$ -- the star formation efficiency \citep{r21}. The return of (enriched) gas to the environment occurs more slowly and is regulated by the stellar lifetimes $\tau_\mathrm{m}$, which in turn depend on both stellar mass and metallicity.

As an ansatz, we assumed that the following relations hold and remain valid throughout the entire process of star formation within the halos:

\begin{equation}
\label{ansatz1}
a_\mathrm{b-H}(t) = 0.7512\, a_\mathrm{b}(t),
\end{equation}
\begin{equation}
\label{ansatz2}
a_\mathrm{b-He}(t) = 0.2487\, a_\mathrm{b}(t),
\end{equation}
\begin{equation}
\label{ansatz3}
a_\mathrm{b-Li}(t) = 4.94\times 10^{-10} a_\mathrm{b}(t).
\end{equation}Once the parameters and functions are characterized, it is possible to write the differential equation for the mass density of the element lithium as
\begin{multline}
\label{lith1}
\frac{d\rho_\mathrm{Li}}{dt} =  \int_{m(t)}^{m_\mathrm{s}}\,[(m - m_\mathrm{r})\,Z_\mathrm{Li}\,(t - \tau_\mathrm{m}) + P_{Z_{\mathrm{Li,m}}}]\, \dot\rho_{\star} (t - \tau_\mathrm{m})\\ \varphi (m)\,dm - Z_\mathrm{Li}\,\dot\rho_{\star}(t) + a_\mathrm{b- Li}(t),
\end{multline}
where the term $(m - m_\mathrm{r})\,Z_\mathrm{Li}\,(t - \tau_\mathrm{m})$ accounts for the amount of lithium incorporated when the star was born, which later returns to the interstellar medium ($m_\mathrm{r}\,Z_\mathrm{Li}\,(t - \tau_\mathrm{m})$ is the part of the lithium retained in the stellar remnant), and the term $Z_\mathrm{Li} = \rho_\mathrm{Li}/\rho_\mathrm{g}$ (where $\rho_\mathrm{g}$ is the total gas density). The $P_{Z_\mathrm{Li,m}}$ parameter is the lithium mass produced by a star of mass $m$, the term $Z_\mathrm{Li}\,\dot\rho_{\star}(t)$ takes into account the removal of part of the lithium element to form a new generation of stars, and $a_\mathrm{b-Li}(t)$ is the primordial lithium incorporated into structures.

The total gas density contained in the term $Z_\mathrm{Li} = \rho_\mathrm{Li}/\rho_\mathrm{g}$ is obtained by solving the equation

\begin{equation}
\label{gas_total}
 \dot\rho_\mathrm{g} = -\frac{d^{2}M_{\star}}{dVdt} + \frac{d^{2}M_\mathrm{ej}}{dVdt} + [a_\mathrm{b-H}(t) + a_\mathrm{b-He}(t)],
\end{equation}
where the first term on the right-hand side describes the removal of gas to form stars, the second term describes the return of gas to the halos through material ejected at the end of the stars' life cycle, and the last term represents the infall of primordial gas.

There are some differences between Eq. (\ref{lith1}) of this study and that used by CMW. Those authors wrote Eq. (\ref{lith1}) for each of the $i$ metals they studied, satisfying the initial condition (at $z=20$) $\rho_\mathrm{g,i} = 0$. Thus, the term $a_\mathrm{b}(t)$ was included through the total gas density at $Z_\mathrm{i} = \rho_\mathrm{g,i}/\rho_\mathrm{g}$. In the present study, primordial lithium is continuously inserted into the structures, an effect described by the last term of Eq. (\ref{lith1}). In turn, the contribution of primordial gas to the total density, according to Eq. (\ref{gas_total}), is mainly due to hydrogen and helium, as represented by the term $[a_\mathrm{b-H}(t) + a_\mathrm{b-He}(t)]$ in Eq. (\ref{gas_total}). With these conditions established, the numerical integration of Eq. (\ref{lith1}) gives the mass density $\rho_\mathrm{Li}$ contained in the structures (DM halos). This allows us to determine the quantity $^{7}\mathrm{Li}/\mathrm{H}$ and compare it with observational data.

In Sect. \ref{sec3} we analyze the role of the $a_\mathrm{b-Li}$ function in the $^{7}\mathrm{Li}/\mathrm{H}$ ratio, as well as the contributions of Pop III and Pop II stars to the results. For comparison with the model, the following observational datasets were selected: 7 extremely metal-poor stars (EMPSs) from \citet{r38}, 41 plateau stars from the sample of \citet{r39}, 49 Gaia-Enceladus sources from the sample of \citet{r40}, the lithium abundance in the Small Magellanic Cloud (SMC) as determined by \citet{r41} and \citet{r42}, and the $^{7}\mathrm{Li}/\mathrm{H}$ abundance in Solar System meteorites presented in \citet{r43}. We also included two extremely iron-poor stars, namely J0023+0307, an extremely dwarf main-sequence (MS) star \citep{r44}, and SMSS J0313--6708, considered one of the oldest stars in the Galaxy \citep{r45}. Finally, we took the primordial limits for the ratio $(^{7}\mathrm{Li}/\mathrm{H})_\mathrm{BBN+CMB}$ from the results of \citet{r5} and \citet{r6}.

\section{Results and discussion}
\label{sec3}

The parameters used for modeling and characterizing the cosmological framework are the same as those used by CMW. These parameters are summarized in Table \ref{tab03}.

\begin{table}[!ht]
\begin{center}
\caption{Cosmological framework parameters.}
\small
\centering
\begin{tabular}{ccccccccc}
\hline
\hline 
$\Omega_{0,\mathrm{m}}$ & $\Omega_{0,\mathrm{b}}$ & $\Omega_{0,\Lambda}$ & $h$ & $z_\mathrm{i}$ & $\sigma_{8}$ & $n_\mathrm{p}$ & $M_\mathrm{min}(M_{\odot})$ \\ 
\hline 
0.279& 0.0463 & 0.721 & 0.7 & 20 & 0.84 & 0.967 & $10^{6}$  \\ 
\hline
\end{tabular} 
\label{tab03}
\end{center}
\tablefoot{
$\Omega_{0,\mathrm{m}}$ is the total matter (baryonic plus DM) density parameter; $\Omega_{0,\mathrm{b}}$ corresponds to the baryonic density parameter; $\Omega_{0,\Lambda}$ is the density parameter associated with the cosmological constant; $h$ is the Hubble constant written as $H_{0}=100\,h\,\mathrm{km}\,\mathrm{s}^{-1}\,\mathrm{Mpc}^{-1}$; $z_\mathrm{i}$ is the redshift at which star formation begins; $\sigma_{8}$ is associated with the normalization of the power spectrum; $n_\mathrm{p}$ is the spectral index of the power spectrum; $M_\mathrm{min}$ corresponds to the lowest mass a DM halo must have to detach from the expansion of the Universe, to collapse, and to virialize, and it is the Jeans mass at recombination (see \citealp{r21})}.
\end{table}

\subsection{The formation of the Spite plateau}
\label{sub31}

\begin{figure}[!ht]
\centering
\includegraphics[width=\hsize]{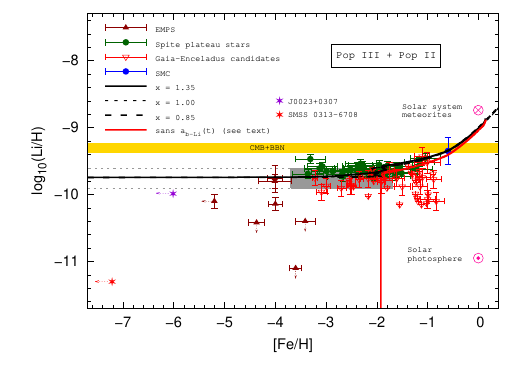}
\caption{Lithium abundance as a function of $\mathrm{[Fe/H]}$. The gray shaded area represents the usual Spite plateau. The primordial infall term in association with the formation of Pop III and Pop II stars causes the Spite plateau to be extended to $\mathrm{[Fe/H]} \lesssim -8.0$.}
 \label{fig01}
\end{figure}

The first result of this investigation is the confirmation that the model described and characterized in the previous section accurately reproduces the lithium abundance observed in the stars distributed across the Spite plateau. As shown in Fig. \ref{fig01}, the gray shaded area represents the typical Spite plateau, extending in the range $-3.7 \leq \mathrm{[Fe/H]} \leq -1.7$. The gray dashed line marks the extension of the Spite plateau up to $\mathrm{[Fe/H]} \sim -8.0$, allowing us to evaluate the behavior of the model for the $^{7}\mathrm{Li/H}$ versus $\mathrm{[Fe/H]}$ ratio. It is important to note that, without adding any new parameters to the CMW model, and by simply applying the ansatz defined by Eqs. (\ref{ansatz1}), (\ref{ansatz2}), and (\ref{ansatz3}), we can confirm that the $^{7}\mathrm{Li}$ abundance predicted by this model fits very well with the usual Spite plateau. Moreover, the model suggests that the Spite plateau extends up to a metallicity of $\mathrm{[Fe/H]} \lesssim -8.0$.

The yellow shaded band indicates the limits for the primordial $^{7}\mathrm{Li}$ abundance determined from the combination of BBN and CMB data. The difference between $(^{7}\mathrm{Li/H})_{\mathrm{BBN+CMB}} \sim 5.0\times 10^{-10}$ and the model's predicted value of $(^{7}\mathrm{Li/H}) \sim 1.8\times 10^{-10}$ is primarily due to the capture of primordial lithium during the birth of Pop III stars within structures formed at $z \sim 20$. The high star formation efficiency $<\varepsilon_{\star}> \sim 0.34$ is responsible for removing primordial gas from the system, keeping the $^{7}\mathrm{Li}$ abundance within the Spite plateau.

Note that the Pop III branch is divided into three classes, as shown in Table \ref{tab01}. Once formed, the stars follow a distribution $\propto m^{-(1+x)}$, where $x$ is the IMF exponent. Stars from the CL08 (low-mass) and HW10 (intermediate-mass) classes of the Pop III branch return some of the primordial lithium they initially incorporated at the end of their lives. During their evolution, stars in these classes also produce lithium, which is partially returned to the system when they reach the end of their life cycle. On the other hand, stars from the HW02 class end their lives as pair-instability supernovae (PISNe), which removes $^{7}\mathrm{Li}$ but does not return it. Additionally, the HW02 group leaves no remnants after the disruptive process is complete.

When the metallicity of the Universe exceeds the threshold $Z = 10^{-6}$, the model assumes that no new Pop III stars can form. However, stars formed within the metallicity range $[0-10^{-6}]$ will continue their evolutionary trajectories as dictated by their lifetimes $\tau_\mathrm{m}$ (see Sect. \ref{sub21}). In other words, the threshold $Z = 10^{-6}$ marks the beginning of Pop II stars' contribution to the chemical enrichment of the Universe, starting with the CL04 group at $Z = 10^{-6}$. The evolutionary paths of the other Pop II branches are described in Sect. \ref{sub21}.

We analyzed three different values for the IMF exponent $x$: 0.85, 1.00, and 1.35 (the Salpeter exponent). The results are only weakly dependent on the value of $x$. Figure \ref{fig01} illustrates the importance of the term $a_\mathrm{b-Li}$ in establishing the Spite plateau when the model includes contributions from both Pop III and Pop II stars.

Excluding the $a_\mathrm{b-Li}$ term from Eq. (\ref{lith1}), as shown by the red curve for $x = 1.35$, results in the $^{7}\mathrm{Li/H}$ ratio reaching the Spite plateau only when $\mathrm{[Fe/H]} > -2.0$. For metallicities above $\mathrm{[Fe/H]} > -2.0$, the curves with (black) and without (red) the $a_\mathrm{b-Li}$ term show similar behavior. Thus, at first glance, the Spite plateau could be seen as a signature of primordial infall. Although the zero-metallicity stars from the CL08 and HW10 branches incorporate the primordial abundances of H, He, and Li, the fraction of $^{7}\mathrm{Li}$ ejected by these stars at the end of their lives is generally smaller than their initial compositions. In contrast, the HW02 branch removes $^{7}\mathrm{Li}$ during the star formation process and does not return any $^{7}\mathrm{Li}$ at the end of these stars' lives. Therefore, the term $a_\mathrm{b-Li}$ is crucial for establishing the Spite plateau at high redshifts (or low metallicities as represented by $\mathrm{[Fe/H]} < -2.0$).

Once the Universe begins to receive contributions from Pop II stars' mass ejections, particularly from the $Z = 10^{-3}$ branch, the $^{7}\mathrm{Li}$ abundance no longer strongly depends on the $a_\mathrm{b-Li}$ term. This occurs when $\mathrm{[Fe/H]} > -2.0$. In Sect. \ref{sub33} we assess this analysis by removing the HW02 branch from the model. Finally, the model accurately reproduces the $^{7}\mathrm{Li}$ abundance in the SMC.

Thus, by incorporating the ansatz (Eqs. \ref{ansatz1}, \ref{ansatz2}, and \ref{ansatz3}) into the CMW model, we can obtain the observed $^{7}\mathrm{Li/H}$ ratio of the Spite plateau and align it with the primordial abundance. Furthermore, as discussed earlier, the Spite plateau is naturally generated and can extend to very low metallicities.

It is important to highlight that Fig. \ref{fig01} shows a large dispersion of the lithium abundance for $\mathrm{[Fe/H]} < 4.0$, which corresponds to the EMPS sample. These seven stars were analyzed by \citet{r38}, and as discussed by those authors, stars with $T_\mathrm{eff}>6000\,K$ and $\mathrm{[Fe/H]}=-4.0$ lie on the Spite plateau, whereas EMPSs with $T_\mathrm{eff} < 6000\,K$ and $\mathrm{[Fe/H]} < -3.5$ always lie below it. According to the authors, this suggests that the $T_\mathrm{eff}$ threshold at which lithium destruction becomes important increases as metallicity decreases. This is likely one of the factors involved in the observed spread relative to the Spite plateau. Other factors likely contribute to this spread as well, such as turbulence and atomic diffusion.

However, the model does not reproduce the lithium abundance observed in Solar System meteorites. This discrepancy may indicate the need to introduce a new source of lithium not accounted for in the CMW model.

\subsection{Lithium production by nova systems}
\label{sub32}

As discussed in \citet{r46}, the detection of lithium in classical nova outbursts suggests that these systems could play a significant role in the production and enrichment of $^{7}\mathrm{Li}$ in galactic systems. In particular, only binary systems formed by stars with $0.8 < M/M_{\odot} < 8$ can develop nova systems capable of producing $^{7}\mathrm{Li}$. As a result, the total mass ($M_{\mathrm{B}}$) of the $^{7}\mathrm{Li}-$producing binary system lies in the range $1.6 < M_{\mathrm{B}}/M_{\odot} < 16$. The distribution function for the primary and secondary stars in the binary system is given by (see \citealp{r47,r46} for details)
\begin{equation}
\label{nova1}
f(\mu) = 2^{1+\gamma}(1+\gamma)\, \mu^{\gamma} \;\;\;\;\;\; (0<\mu\leq 1/2),
\end{equation}
where $\mu = M_{\mathrm{sec}}/M_{\mathrm{B}}$, with $M_{\mathrm{sec}}$ being the mass of the secondary star, and $\gamma = 2$.

Additionally, for $^{7}\mathrm{Li}$ production to occur in a nova system, the primary star must evolve into a white dwarf and accumulate enough material from the secondary star to reach the ignition conditions for the nova outburst. This implies that binaries with stars of the same mass cannot develop nova systems. The lifetimes of the primary and secondary stars must satisfy the condition $\tau_{\mathrm{sec}} - \tau_{\mathrm{prim}} \geq \tau_{\mathrm{nova}}$, where $\tau_{\mathrm{nova}}$ is the delay time.

The $^{7}\mathrm{Li}$ production in nova systems can be determined by

\begin{multline}
\label{nova2}
 \dot\rho_{\mathrm{N-Li}} = \mathcal{N}_{\mathrm{Li}} \int_{M_\mathrm{Bm}}^{M_{\mathrm{BM}}} \phi(M_{\mathrm{B}}) \\ \left \{\int_{\mu_{\mathrm{m_{min}}}}^{1/2} f(\mu)\, \dot\rho_{\star}(t-\tau_{\mathrm{sec}})\,d\mu \right \} dM_{\mathrm{B}}\, \mathcal{Y}_{\mathrm{N-Li}},
\end{multline}
where $\mu_{\mathrm{m_{min}}}$ is the minimum mass fraction contributing to the nova rate at time $t$, $\phi(M_{\mathrm{B}})$ is the IMF for the total mass of the binary system ($\phi(M_{\mathrm{B}}) \propto M_{\mathrm{B}}^{-(1+x)}$), $\dot{\rho}_{\star}(t-\tau_{\mathrm{sec}})$ is the CSFR at the time when the secondary star is formed, $\mathcal{Y}_{\mathrm{N-Li}}$ is the amount of $^{7}\mathrm{Li}$ produced in each nova event, and $\mathcal{N}_{\mathrm{Li}}$ is a fixed parameter to reproduce the rate of nova outbursts in the Galaxy at the present time.

The nova rate is determined following the formalism presented in \citet{r48}. As discussed by these authors (and also by \citealp{r46}), the nova outburst rate is related to the white dwarf formation rate in binary systems. Each nova system can produce an average of $10^{4}$ nova outbursts. These points are considered in determining the parameter $\mathcal{N}_{\mathrm{Li}}$, which is adjusted to reproduce a nova rate of $50\,\mathrm{yr}^{-1}$, in agreement with recent estimates of the Galactic nova rate, yielding a value of $50^{+31}_{-23}\,\mathrm{yr}^{-1}$ (\citealp{r49}). We determined the yield parameter $\mathcal{Y}_{\mathrm{N-Li}}$ to match the observed $^{7}\mathrm{Li}$ abundance in meteorites. Finally, we adopted $\tau_{\mathrm{nova}} = 1\,{\mathrm{Gyr}}$ for the delay time (\citealp{r48,r46}).

\begin{figure}[!ht]
\centering
\includegraphics[width=\hsize]{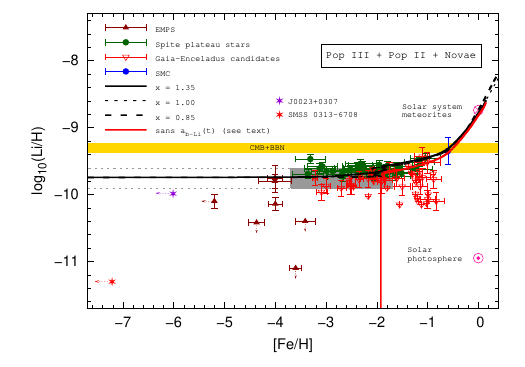}
\caption{Lithium abundance as a function of $\mathrm{[Fe/H]}$. The figure shows the contribution of nova binary systems to the lithium abundance when $\mathrm{[Fe/H]} > -1.0$.
}
 \label{fig02}
\end{figure}

In Fig. \ref{fig02} we present the contribution of nova outbursts to the lithium yield by incorporating Eq. (\ref{nova2}) into the right-hand side of Eq. (\ref{lith1}). Nova systems become effective at producing $^{7}\mathrm{Li}$ only for $\mathrm{[Fe/H]} > -1.0$. This delay is due to the time required for the primary star to evolve into a white dwarf, as well as the time necessary for the white dwarf to cool enough for a nova outburst to occur.

The parameters $\mathcal{N}_{\mathrm{Li}}$ and $\mathcal{Y}_{\mathrm{N-Li}}$ required to produce a Galactic nova rate of $50\,\mathrm{yr}^{-1}$ and return the lithium abundance observed in meteorites are shown in Table \ref{tab04}. We used $\tau_{\mathrm{nova}} = 1\,{\mathrm{Gyr}}$ in this model, consistent with the fiducial model of \citet{r46}. It is important to highlight some differences between this work and that of \citet{r46}. While they explored the values 0, 1, 2, and $5\,{\mathrm{Gyr}}$ for $\tau_{\mathrm{nova}}$, they considered $1\,{\mathrm{Gyr}}$ as the best value for the delay time. Another key difference concerns the parameter $\mathcal{N}_{\mathrm{Li}}$. They found that a value of $0.03$ provided a nova burst rate compatible with $20-30\,{\mathrm{yr}}^{-1}$. However, we adopted the most recent value (see \citealp{r49}) of $50^{+31}_{-23}\,\mathrm{yr}^{-1}$, which is approximately $2-3$ times larger than the rate used by \citet{r46}.
\begin{table}[!ht]
\begin{center}
\caption{Parameters used in Eq. (\ref{nova2}) to determine the lithium yield of nova binary systems.}
\small
\centering
\begin{tabular}{ccccccccc}
\hline
\hline 
$x$ & $\mathcal{N}_{\mathrm{Li}}$ & $\mathcal{Y}_{\mathrm{N-Li}}\,(M_{\odot})$ \\ 
\hline 
0.85 & 0.301 & $3.0 \times 10^{-4}$  \\ 
\hline
1.00 & 0.170 & $2.0 \times 10^{-4}$  \\
\hline
1.35 & 0.066 & $5.4 \times 10^{-5}$ \\
\hline
\end{tabular} 
\label{tab04}
\end{center}
\end{table}

In our model, for a nova outburst with $x=1.35$, we obtain $\mathcal{N}_{\mathrm{Li}} = 0.066$, a result consistent with the analogous factor of \citet{r46}. Finally, \citet{r46} found that the lithium abundance in meteorites could be matched with $\mathcal{Y}_{\mathrm{N-Li}} = 4.14 \times 10^{-5} M_{\odot}$, while we obtain agreement with $\mathcal{Y}_{\mathrm{N-Li}} = 5.4 \times 10^{-5} M_{\odot}$ (for $x=1.35$). Assuming each nova system undergoes $\sim 10^{4}$ outbursts during its lifetime, the required lithium production per nova outburst is approximately $5.4 \times 10^{-9} M_{\odot}$ in our model.

\subsection{Lithium from the spallation of CNO by Galactic cosmic rays}
\label{sub33a}

Type II supernovae are natural sites for the generation and acceleration of high-energy cosmic rays, which can, in turn, produce elements such as lithium, beryllium, and boron through spallation of CNO nuclei in the Galactic environment (\citealp{r54}; \citealp{r55}; \citealp{r19}). As discussed by \citet{r55}, GCRs may play an important role in $^{7}\mathrm{Li}$ production for $\mathrm{[Fe/H]} > -3.0$, possibly complementing the contribution of novae. To include GCRs in our model, we obtained a fit consistent with the curve shown in Fig. 1 of \citet{r55} for $\mathrm{[Fe/H]} > -3.0$. Combining GCRs and novae, it is possible to reach the meteoritic $^{7}\mathrm{Li}$ abundance with $9.3\%$ from GCRs and $90.7\%$ from novae. This allows for a $16.7\%$ reduction in the total lithium yield in the model with IMF slope $x = 1.35$ (fiducial model from Table \ref{tab04}), resulting in $\mathcal{Y}_{\mathrm{N-Li}} = 4.5\times 10^{-5}\,M_{\odot}$. This value is closer to the estimate of \citet{r46}. The yield reduction factor of $16.7\%$ also applies to the other models in Table \ref{tab04}, particularly with $\mathcal{Y}_{\mathrm{N-Li}} = 1.67\times 10^{-4}\,M_{\odot}$ for $x = 1.00$ and $2.50\times 10^{-4}\,M_{\odot}$ for $x = 0.85$.

\subsection{Lithium production without the HW02 branch of Population III stars}
\label{sub33}

Between zero and $10^{-6}$ metallicity, Pop III stars from branches HW02 and HW10 contribute predominantly to the chemical enrichment of the environment, with HW02 not producing lithium but instead removing it from the system. Thus, the lithium content in the star-forming environment is slightly higher -- specifically, 1.002 (or $\sim 0.2\%$ higher) -- when only the HW10 branch is present. The presence or absence of the HW02 branch has a more significant impact on the amount of iron. Without HW02, iron injection into the star-forming environment occurs more slowly. The amount of iron injected jointly by HW02 and HW10 is approximately $3.2$ times greater than that injected by HW10 alone during the metallicity phase from zero to $10^{-6}$. Therefore, removing HW02 Pop III stars slightly increases lithium but substantially reduces iron in the metallicity range between zero and $10^{-6}$. Since the reference metallicity of $10^{-6}$ marks the beginning of Pop II star formation, the subsequent chemical evolution in the star-forming environment becomes linked to the contribution of Pop III stars.

In particular, a Pop III star with an initial mass of $\sim 200\,M_{\odot}$ (HW02 branch) returns $\sim 100\,M_{\odot}$ in metals and $\sim 10\,M_{\odot}$ in iron (\citealp{r32}). PISNe end their lives without leaving any remnants due to a complete disruptive process. Given the characteristics of these stars, the results for $\mathrm{[Fe/H]}$ depend on the presence of the PISN branch. Moreover, these stars remove $^{7}\mathrm{Li}$ from the environment and do not return it at the end of their lives.

\begin{figure}[!ht]
\centering
\includegraphics[width=\hsize]{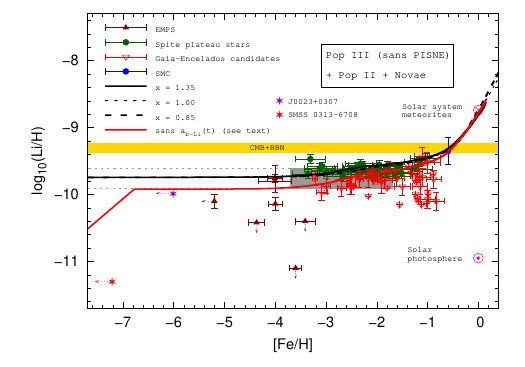}
\caption{Lithium abundance versus $\mathrm{[Fe/H]}$. In this case, only the CL08 (low-mass) and HW10 (intermediate-mass) branches of Pop III stars are included.
}
 \label{fig03}
\end{figure}

On the other hand, a Pop III star from the HW10 branch with $\sim 100\,M_{\odot}$ returns $\sim 9\,M_{\odot}$ in metals, $\sim 10^{-12}\,M_{\odot}$ in iron, and $\sim 10^{-6}\,M_{\odot}$ in lithium, assuming the standard model with an energy explosion of $1.2\,\mathrm{B}$ (see \citealp{r33} for details). As a result, the $\mathrm{[Fe/H]}$ versus redshift relationship depends on the presence (or absence) of HW02 branch stars. Figure \ref{fig03} illustrates the behavior of the $\mathrm{[Fe/H]}$ versus $^{7}\mathrm{Li/H}$ relationship without the PISN branch. A comparison of the results shown in Figs. \ref{fig02} and \ref{fig03}, with $a_\mathrm{b-Li}$, reveals virtually no difference. There is only a slight increase in the amplitude of $^{7}\mathrm{Li/H}$ for the model sans PISNe (Fig. \ref{fig03}) compared to the model with PISNe (Fig. \ref{fig02}). This happens because stars from the HW10 and CL08 branches synthesize $^{7}\mathrm{Li}$, while stars from the PISN branch (HW02) only remove this element from the environment.

The main difference between the models in Figs. \ref{fig02} and \ref{fig03} is observed when we compare the red curves for the models without the primordial infall term. In the model with PISNe, the transition from Pop III to Pop II (with metallicity $Z=10^{-6}$) occurs when $\mathrm{[Fe/H]} \sim -2.3$. This transition happens very quickly due to the large amount of iron injected by the HW02 stars. Only when $\mathrm{[Fe/H]} > -2.3$ do the stars of the HW02 branch stop forming and removing $^{7}\mathrm{Li}$ from the system. During this phase, lithium enrichment in the structures primarily comes from synthesis by stars in the HW10 branch, while stars in the HW02 branch are responsible for determining $\mathrm{[Fe/H]}$.

For the sans PISNe model (Fig. \ref{fig03}), during the phase characterized by the chemical enrichment of the Universe from zero metallicity up to $Z=10^{-6}$, stars from the HW10 branch are responsible for enriching both $^{7}\mathrm{Li}$ and $\mathrm{[Fe/H]}$. The transition to Pop II is slower without the HW02 branch, occurring when $\mathrm{[Fe/H]} \sim -3.1$ (corresponding to $Z=10^{-6}$). Although Pop III stars no longer form when $Z>10^{-6}$, the lower-mass stars from the HW10 branch, along with those from the CL08 branch, continue to contribute to $^{7}\mathrm{Li}$ injection into the system up to $\mathrm{[Fe/H]} \sim -0.84$, at which point they end their lives, and lithium synthesis now depends solely on Pop II stars from the various branches listed in Table \ref{tab02}.

Despite the fact that the sans PISNe model achieves a $^{7}\mathrm{Li/H} \sim 1.20\times 10^{-10}$ without the contribution of the infall term, this value represents the lower limit of the usual Spite plateau. Only by including the primordial infall term can $^{7}\mathrm{Li/H} \sim 1.81\times 10^{-10}$ be maintained, which is closer to the central value of the Spite plateau. From Fig. \ref{fig03}, we can see that without the term $a_\mathrm{b-Li}$, in the sans PISNe model, the extended Spite plateau breaks down for $\mathrm{[Fe/H]} < -6.7$, whereas with the primordial infall term (black lines in Fig. \ref{fig03}), the Spite plateau is maintained up to $\mathrm{[Fe/H]} \lesssim -8.0$.

\subsection{Lithium without the contribution of the Pop III stars}
\label{sub34}

\begin{figure}[!ht]
\centering
\includegraphics[width=\hsize]{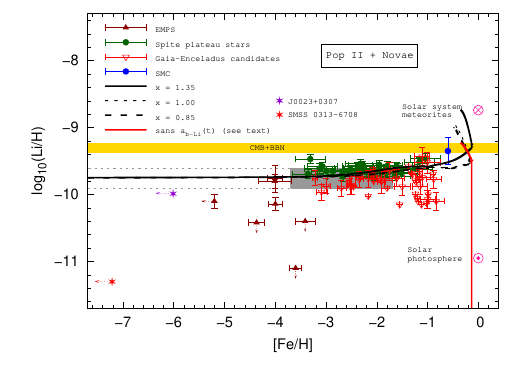}
\caption{Lithium abundance versus $\mathrm{[Fe/H]}$ considering the contribution of Pop II stars and nova systems working together with GCRs. The Spite plateau is established by the primordial infall term. Even with the infall term, only Pop II stars, novae, and GCRs do not return the lithium abundance inferred for Solar System meteorites (see the main text).
}
 \label{fig04}
\end{figure}

Figure \ref{fig04} shows the behavior of $^{7}\mathrm{Li/H}$ considering only Pop II stars. In this case, similar to the study by CMW, it is assumed that zero-metallicity stars have yields identical to Pop II stars with $Z=10^{-6}$ (CL04 branch, as indicated in Table \ref{tab02}). These stars are responsible for synthesizing lithium, iron, and other elements, chemically enriching the Universe up to $Z=10^{-4}$, when stars from the subsequent branches such as K10, D14b, and CL04 begin to form, as described in Sect. \ref{sub21}.

Stars from the CL04 branch (with $Z=10^{-6}$) return $3.5\,{M_\odot}$ in metals, $10^{-1}M_{\odot}$ in iron, and $\sim 3.0\times 10^{-11}M_{\odot}$ in lithium at the end of their lives. The cycle characterized by the formation of these stars continues until the metallicity of $Z=10^{-4}$ is reached, at which point the formation of stars with this metallicity begins. Although the iron production of stars from the CL04 (with $Z=10^{-6}$) and HW10 branches is similar, Pop II stars with $Z=10^{-6}$ produce one-third of the total amount of metals and much less lithium compared to HW10 stars. As a result, the values of $\mathrm{[Fe/H]}$ and $^{7}\mathrm{Li/H}$ increase much more slowly over time.

It can be seen in Fig. \ref{fig04} that the Spite plateau is established through the primordial infall term ($a_\mathrm{b-Li}$). The lithium production by Pop II stars with $Z=10^{-6}$ is $\sim 10^{5}$ times lower than the average yield of Pop III stars from the HW10 branch. The next class of Pop II stars with $Z=10^{-4}$ contributes very weakly to the production of $^{7}\mathrm{Li}$. Most stars in the $1-7\,M_{\odot}$ range have negative yields, contributing to the removal of $^{7}\mathrm{Li}$ from the system. Stars in the $7.5-35\,M_{\odot}$ range, on average, synthesize five times less lithium than stars in the $Z=10^{-6}$ class.

The stars of the next metallicity class ($Z=10^{-3}$) return an average of $\sim 10^{-8}M_{\odot}$ of $^{7}\mathrm{Li}$, which is greater than the contribution from stars in the less enriched Pop II classes. The contribution from novae occurs simultaneously with that of stars from the $Z=10^{-3}$ class, increasing the value of $^{7}\mathrm{Li/H}$ when $-0.13 > \mathrm{[Fe/H]} > -0.70$. Pop II stars from subsequent classes (i.e., $Z=4\times 10^{-3}$ and $Z=8\times 10^{-3}$) mostly have negative yields, leading to the incorporation and destruction of $^{7}\mathrm{Li}$ in the environment. During this phase, the enrichment of lithium in the structures comes from the infall of primordial gas and nova systems together GCRs. Additionally, stars from these two metallicity classes produce almost no iron. This explains the behavior of the $^{7}\mathrm{Li/H}$ versus $\mathrm{[Fe/H]}$ ratio, where the abundance of lithium increases while the availability of iron in the environment decreases.

Although the inclusion of the term $a_\mathrm{b-Li}$ produces the Spite plateau, it is important to emphasize that by considering only the yields from Pop II stars and novae working together with GCRs, the relation between $^{7}\mathrm{Li/H}$ and $\mathrm{[Fe/H]}$ observed in Solar System meteorites cannot be recovered. This is due to the characteristics of stars with metallicities of $4\times 10^{-3}$ and $8\times 10^{-3}$, as discussed earlier. On the other hand, the removal of the term $a_\mathrm{b-Li}$ (red curve in Fig. \ref{fig04}) does not provide a physically adequate model for the $^{7}\mathrm{Li/H}$ versus $\mathrm{[Fe/H]}$ relation.

It is crucial to emphasize, as highlighted in Sect. \ref{sub21}, that the physical basis for the Pop II yields used in this study includes the MONSTAR code for stellar evolution, OPAL opacities, and compatible mass loss models, providing a common physical foundation for chemical modeling across stellar populations from metallicity $10^{-6}$ to $2\times 10^{-2}$ \citep{r21}. In particular, MONSTAR treats the transport of chemical elements by convection only, not accounting for diffusion or turbulence. On the other hand, codes such as MESA+NuGrid (see, e.g., \citealp{r59}) incorporate microscopic diffusion, rotation, and magnetic fields.

Comparison between the $^{7}\mathrm{Li}$ yields from this work and those from MESA+NuGrid is not direct for the same stellar mass and metallicity ranges presented in Table \ref{tab02}. Within the possible comparisons for Pop II stars at specific metallicities $Z \sim 10^{-6}$ and $10^{-4}$, the present work shows higher $^{7}\mathrm{Li}$ yields for the mass range $5$ to $25\,M_{\odot}$, while above $25$ to $60\,M_{\odot}$ MESA+NuGrid with diffusion gives higher lithium yields. Integrating the total $^{7}\mathrm{Li}$ yield from $5$ to $60\,M_{\odot}$, this work produces approximately $1.5$ times more lithium than the diffusion and rotation inclusive models. This result agrees with the analysis by \citet{r53}. Thus, within the possible comparisons to be made between MONSTAR and MESA+NuGrid, the removal of the $a_\mathrm{b-Li}$ term would not produce a scenario capable of generating the Spite plateau with only Pop II stars.

\subsection{J0023+0307 and SMSS J0313--6708}
\label{sub35}

While this work does not aim to evaluate the $^{7}\mathrm{Li/H}$ versus $\mathrm{[Fe/H]}$ ratio for individual objects, it is still worthwhile to make some observations about J0023+0307 and SMSS J0313--6708 in light of the model presented here. The inferred abundances $\mathrm{[Fe/H]} \lesssim -6.1$ ($\mathrm{[Fe/H]} \lesssim -7.1$ for SMSS J0313--6708) classify these objects as extremely iron-poor stars.

J0023+0307 is a dwarf star with $A\mathrm{(Li)} = 2.02$ \footnote{$A\mathrm{(Li)} = \log_{10}\mathrm{(Li/H)} + 12$.} \citep{r44}, placing it at the lower limit of the Spite plateau. Since no iron lines were detected in the spectrum of this star, the upper limit of $\mathrm{[Fe/H]} = -6.1$ was inferred from the resonant $\mathrm{Ca \,II}$ lines, assuming $\mathrm{[Ca/Fe]} \geq 0.40$ \citep{r44}. The chemical abundance analysis of J0023+0307 by \citet{r51}, using supernova yield models from HW10, suggested that J0023+0307 formed in a re-collapsed minihalo. Specifically, using nonlocal thermodynamic equilibrium abundances, the authors identified four best-fit models from the entire HW10 database. Using the StarFit algorithm \footnote{www.2sn.org/starfit}, we obtained the same set of models as \citet{r51}. The model with the lowest $\chi^{2}$ corresponds to a star with a mass of $50\,M_{\odot}$, an explosion energy of $3.0\,\mathrm{B}$, a mixing parameter of $3.98\times 10^{-3}$, and a remnant black hole mass of $\sim 3.6\,M_{\odot}$.

The condition $\mathrm{[Fe/H]} = -6.1$ is met in our model when the total metallicity of the Universe is $Z=2.3\times 10^{-10}$ in the scenario that includes PISNe ($Z=9.8\times 10^{-10}$ without PISNe) and assuming an IMF exponent of $x=1.35$. J0023+0307 would have formed approximately $4.4\times 10^{5}\,\mathrm{yr}$ after the explosion of the first PISN or $\sim 1.3\times 10^{6}\,\mathrm{yr}$ after the explosion of the first star in the HW10 branch (model without PISNe).

The statistical analysis by \citet{r51} suggests that the inferred dilutions of hydrogen masses ($M_\mathrm{dil}$) from their fits range from $10^{4.5} - 10^{5.5}M_{\odot}$, which is consistent with the idea that J0023+0307 formed in a re-collapsed minihalo. Since the present study, along with the CMW model, incorporates the structure of the unified model developed by \citet{r26}, it is also plausible that J0023+0307 could have formed in a subhalo (a globular cluster-like substructure) with a mass of $\sim 4.6\times 10^{4}M_{\odot}$, contained within one of the first halos with a mass of $10^{6}M_{\odot}$. While these scenarios are distinct, there is a similarity between our findings and the proposal of a re-collapsed minihalo by \citet{r51} to explain the chemical abundances of this extremely iron-poor star.

SMSS J0313--6708 is a red giant with $\mathrm{[Fe/H]} < -7.1$ and $A\mathrm{(Li)} = 0.7$ \citep{r45,r52}. In particular, \citet{r45} compared the abundances of SMSS J0313--6708 with the yields of Pop III supernovae and obtained a good fit to a progenitor star with a mass of $60\,M_{\odot}$ and primordial composition. Using the HW10 database and the StarFit algorithm in this study, we reached a similar conclusion: a progenitor star with $60\,M_{\odot}$, an explosion energy of $1.8\,\mathrm{B}$, a mixing parameter of $1.58\times 10^{-3}$, and a remnant black hole mass of $25\,M_{\odot}$.

The condition $\mathrm{[Fe/H]} = -7.1$ is satisfied in our model when the total metallicity of the Universe is $Z=2.3\times 10^{-11}$ in the PISN scenario ($Z=2.3\times 10^{-10}$ without PISNe). The $\mathrm{[Fe/H]} = -7.1$ abundance is reached approximately $2.2\times 10^{5}\mathrm{yr}$ after the first PISN event or $\sim 4.4\times 10^{5}\mathrm{yr}$ after the explosion of the first star in the HW10 branch (model without PISNe).

As SMSS J0313--6708 is in the red giant phase, its lithium abundance must have decreased during its evolution from the MS to the upper red giant branch (RGB), due to the destruction of lithium in the convective envelope during the RGB phase. As discussed by \citet{r52}, it is possible that the lithium abundance was depleted by approximately $1.3\,\mathrm{dex}$, which would imply that the star initially had $A\mathrm{(Li)_{MS}} \sim 2.0$, placing it near the lower boundary of the Spite plateau.

\section{Conclusions}
\label{final}

In this work we address the cosmological lithium problem by employing a chemical model that incorporates yields from stars with metallicities ranging from zero to solar. The present scenario follows the unified CSFR-LSF model proposed by \citet{r26}, which consistently derives Larson's law, among other results. The cosmological structure formation framework used here mirrors that of \citet{r24}, \citet{r26}, and \citet{r21} and is based on the formalism originally developed by \citet{r22} to describe the formation and evolution of large-scale structures in the Universe.

The results of this work show that the initial assumption introduced by the ansatz, expressed in Eqs. (\ref{ansatz1}), (\ref{ansatz2}), and (\ref{ansatz3}), is consistent with explaining the lithium abundance over cosmic time. In particular, the introduction of the ansatz shows that the so-called cosmological lithium problem can be addressed using a standard chemical evolution model coupled to the hierarchical structure formation scenario. Additionally, the model presented in this work does not introduce any additional parameters to the scenario described by \citet{r26} and \citet{r21}, thereby preserving all the results obtained by those authors. This strengthens the interpretation that the ansatz considered here is valid throughout the star formation process.

Additionally, diffusion processes, which reduce the lithium abundance, are evaluated in studies that complement the scenario proposed here. Using the Spite plateau reference established in this work, these processes could decrease lithium abundances by a factor of roughly 1.6 to 2.0, as determined by \citet{r53}. This reduction might partly account for the dispersion observed in the lithium abundance versus $\mathrm{[Fe/H]}$ among the samples presented in Figs. \ref{fig01} through \ref{fig04}.

Our main conclusions are summarized as follows:
\begin{itemize}

\item There is no so-called cosmological lithium problem, i.e., there is no discrepancy between the primordial abundance of $^{7}\mathrm{Li}$ determined from combined BBN+CMB predictions and the inferred abundance in metal-poor stars near the MS.

\item The use of the ansatz enables the generation of the Spite plateau for the $^{7}\mathrm{Li/H}$ versus $\mathrm{[Fe/H]}$ relation, in agreement with the BBN+CMB predictions and inferred abundances for several samples: EMPSs \citep{r38}, plateau stars \citep{r39}, Gaia-Enceladus \citep{r40}, and the SMC \citep{r41,r42}. The term $a_\mathrm{b-Li}$, together with processes associated with stellar formation and the yields of Pop III and Pop II stars, produces an extended Spite plateau within the range $-8.0\lesssim \mathrm{[Fe/H]} \lesssim -2.0$. The inclusion of $a_\mathrm{b-Li}$ allows us to obtain the Spite plateau with $^{7}\mathrm{Li/H} \sim 1.81\times 10^{-10}$, which is consistent with the result found by \citet{r6} and references therein.

\item The enrichment of $^{7}\mathrm{Li}$ over cosmic time is weakly dependent on the IMF. However, $x=1.35$ is the exponent that best brings the parameters of our scenario in line with the fiducial parameters used by \citet{r46} in their study of nova binary systems. In our model, the abundance of $^{7}\mathrm{Li}$ in meteorites is obtained from the combined contribution of novae and GCRs.

\item J0023+0307 could have formed $4.4\times 10^{5}\mathrm{yr}$ after the first PISN explosion or $1.3\times 10^{6}\mathrm{yr}$ after the first star of the HW10 branch (model without PISNe). The results in Sect. \ref{sub35} are similar to the analysis performed by \citet{r51}, particularly their suggestion that the hydrogen mass dilutions $(M_\mathrm{dil})$ inferred from their fits could indicate that J0023+0307 formed in a re-collapsed minihalo. This idea is consistent with the results of \citet{r26}, who propose that part of the star formation at high redshifts occurred in DM subhalos with masses similar to those of globular clusters. In this study, J0023+0307 could have formed in a subhalo-like structure of $\sim 4.6\times 10^{4}M_{\odot}$, contained in one of the first halos formed with a mass of $10^{6}M_{\odot}$ at redshift 20.

\item SMSS J0313--6708 could have formed $2.2\times 10^{5}\mathrm{yr}$ after the first PISN event or $4.4\times 10^{5}\mathrm{yr}$ after the explosion of the first star in the HW10 branch (model without PISNe). As SMSS J0313--6708 is in the red giant phase, its lithium abundance must have decreased during its evolution from the MS to the RGB. As discussed by \citet{r52}, it is possible that this star initially had $A\mathrm{(Li)_{MS}} \sim 2.0$ and was depleted by $1.3\,\mathrm{dex}$ during the RGB phase. Therefore, SMSS J0313--6708 could have initially had a lithium abundance close to the Spite plateau.

\end{itemize}

Although this work demonstrates that the Spite plateau can be generated by three different scenarios, the formation of large-scale structures in the Universe, as described by the Press-Schechter-like formalism, suggests that the $a_\mathrm{b-Li}$ term cannot be omitted. The DM halos that decouple from the expansion of the Universe, collapse, and virialize are immersed in the primordial environment, which consists of elements synthesized during primordial nucleosynthesis. It is in this environment that the potential wells of the halos create conditions for baryonic matter to flow into the structures, agglomerate, and form the first stars. Thus, the physically appropriate scenarios are those that include the $a_\mathrm{b-Li}$ term, operating together with Pop III stars in the low-metallicity regime.

The remaining question to resolve is whether the HW02 branch could have contributed to the chemical enrichment of the Universe. Although the analyses performed here for J0023+0307 and SMSS J0313--6708 support the results of other authors, particularly that the chemical abundances of these stars are compatible with progenitors of the HW10 branch and not the HW02 branch, it is possible that, if part of the star formation at high redshifts occurred in subhalos (see \citealp{r26}) or in re-collapsed minihalos \citep{r51}, the contribution of PISNe to the establishment of the extended Spite plateau would not be ruled out.

Finally, the cosmological lithium problem can be resolved without any changes to standard BBN. The hierarchical structure formation scenario, as described by \citet{r22}, with the appropriate characterization of the CSFR \citep{r23,r24,r26,r27} and yields from Pop III and II stars \citep{r21}, along with Schmidt's law for star formation \citep{r28}, the Salpeter IMF \citep{r29}, nova systems \citep{r47,r48,r49,r46}, and GCRs \citep{r54,r55,r19},   produces a consistent scenario for the observed $^{7}\mathrm{Li}$ abundance over cosmic time.


\bibliography{bibtex} 
\bibliographystyle{aa} 

\end{document}